# Robust PCA Study of TEM Images of Gold Nanospheres: Stable and Dynamic Closed Knot-Like Phase-Space Loops

M. F. Yilmaz[1], B. Karlik[2]

[1]Sheriff LTF LLC, Fremont, CA, USA

[2]Neurosurgical Simulation and Artificial Intelligence Learning Center, Montreal Neurological Institute, McGill University, H3A2B4, Montreal, QC, Canada

**ABSTRACT**

In this study, we apply Robust PCA to transmission electron microscopy (TEM) images of gold nanoparticles (GNPs) to decompose their geometrical, topological, and dynamical organization into static and dynamic knot-like phase-space loops. Dynamic loops show higher crossing numbers and longer rope lengths, while static loops show higher mean curvature. Although most Alexander determinants were $\Delta(t)=1$ with isolated values of $\Delta(t)=3$ at smaller nanoparticle diameters, implying that geometrical complexity does not require topological complexity. The different writhe signs indicate opposite handedness and distinct phase-space geometries in static and dynamic cases. For both static and dynamic knot-like loops, the graphs show an inverse relationship between mean curvature $\langle\kappa\rangle$ and mean torsion $\langle\tau\rangle$. Flatter trajectories occur at high curvature, and more torsional twisting occurs at low curvature. The small changes in Moffatt–Ricca helicities ($\sim 10^{-14} << 0$) are due to circulation-dependent scaling, yet Moffatt–Ricca helicities still follow the same self-linking trend in both static and dynamic configurations. As nanoparticle diameter increases, the dynamic knot-like loops show stronger Kelvin-like spectral localization and smaller minimum self-approach distances, indicating more concentrated spatial modes and tighter geometric folding. At the same time, the decreasing CTEI suggests reduced global phase-space organization. These contrasting trends indicate that the geometrical organization tends to be less global and coherent, and more localized and mode-specific. These results show that the PCA-reconstructed phase space could serve as a mathematical analog to Kelvin's aether, because all disturbances exhibit similar topological configurations with varying degrees of geometric difference.

*Corresponding Author: fthyilmaz4206@gmail.com*

## 1. Introduction

Knotted and linked objects are among the most basic objects in nature. They appear in DNA, proteins, synthetic polymers, etc., and constructing knotted nanoobjects is one of the most active areas of modern chemistry and materials science [1,2]. In nanophotonics, a knotted loop has two meanings. First, the metal is tied in a knot, such as helices, linked rings, torus knots, Möbius-type resonators, etc. These structures show strong circular dichroism and optical rotatory dispersion, many orders higher than those of chiral molecules [3–6], and also exhibit high-Q, non-radiating toroidal and anapole modes [7]. In the second type, the metal is planar, but the field is knotted. Skyrmion and meron lattices, which are topologically protected and feature deep subwavelength features, are created by the interference of surface plasmon polaritons and have been directly imaged by near-field and photoemission microscopy [8–11]. These papers follow the trend to create isolated vortex knots in a free-space light beam [12]. Knotted plasmonic matter and knotted plasmonic fields are expected to have many applications in enantioselective sensing and chiral photodetectors, high-Q non-radiating resonators, topologically robust encoding, and on-chip transport of optical information [5,8,13].

The unique properties of these plasmonic nanostructures offer a powerful method for generating and sustaining knotted optical fields. Due to the LSPR phenomenon, a perfectly metallic nanosphere (achievable only in the Rayleigh limit) focuses EM energy into sub-diffraction-volume regions on its surface. As a result, the plasmonic nanostructure produces much higher spatial field gradients than free-space optics can [14,15]. These spatial gradients enable the generation and sustainment of knotted near-field structures that paraxial or weakly focused optical fields cannot access. TEM and STEM are preferred for resolving the atomic structure, size distribution, and surface morphology of AuNPs. During high-resolution TEM characterization of AuNPs, researchers often observe anomalous structures such as knotted dislocation loops, tangled networks of partial dislocations, and ring-like contrasts in the cores of nanospheres [16,17,18, and 19]. It's important to determine whether these structures arise from inherent crystallographic defects, plasmonic-related environments, or electron-beam damage.

Decoding such highly spatially layered and interlaced field gradients is a major computational challenge, and pattern recognition and Artificial Intelligence technologies play a key role. For example, Karlik and Yilmaz et al. (2019) used hybrid machine-learning architectures to efficiently decode the effects of such nanochain systems [20]. One could show that using PCA on the UV Absorption spectra of a gold sphere nanochain dataset can reveal hidden spatial structures, attractors, and other patterns in the space that are intrinsic to the complex vector field's dynamical evolution. LDA of the nanochain UV spectral data, standing-wave plasmonic oscillations, and their 3D plots would let us see the effects of quantum confinement. Using cutting-edge pattern recognition and deep learning technologies, we could develop a highly useful, general approach to visualize these hidden topological networks and geometric structures. In addition, by building sophisticated data-driven mathematical models, such as RPCA, we could directly parametrize and localize spatial and temporal information in the multi-dimensional Hilbert space [21].

We propose a comprehensive comparative study of knotted plasmonic nanosphere loops under static and dynamic excitation as a function of nanosphere size. To this end, we use 3D simulation data of plasmonic nanospheres and perform an extensive analysis of knot-theoretic, differential-geometric, and vortex-dynamical metrics in these two excitation regimes. Our main idea is that knot invariants correspond naturally to the Moffatt-Ricca theory of vortex helicity. From the Călugăreanu-White-Fuller theorem on the self-linking number being the sum of writhe and twist of a knot, we have a correspondence between the self-linking number of a knotted filament and the Moffatt-Ricca helicity [22,23,24,25]. This natural correspondence can describe the dynamics of plasmonic knots. Our paper is organized as follows: Section 2 presents experimental and theoretical techniques for PCA. Section 3 presents the theory behind plasmonic knot formation, knot invariants, and vortex dynamics, and how we integrate them. Section 4 presents the findings on knot invariants and vortex diagnostics. Section 5 presents our conclusions.

## 2. METHODS

### 2.1 Experiments

We procured hydrogen tetrachloroaurate(III) trihydrate ($HAuCl_4 \cdot 3H_2O$, gold salt), trisodium citrate (reducing agent), and sodium chloride from Aldrich and used them as received. We obtained gold nanoparticles (AuNPs) of 10, 30, 50, and 70 nm (Fig. 1) from Nanocomposix, Inc., and prepared several aqueous AuNP solutions using distilled water. We synthesized Au nanoparticles using the Turkevich, Frens, and Jana reduction methods. With continuous stirring in an ice bath, we added 0.3 mL of 0.1 M sodium borohydride ($NaBH_4$). We confirmed particle formation by observing a color change in the solution. To prepare AuNPs 7 nm and larger, we prepared aqueous solutions of 0.254 mM $HAuCl_4$ and 38.8 mM trisodium citrate. 50 mL of $HAuCl_4$ solution was kept for boiling, and after 5 minutes of boiling, the calculated volume (0.4 mL, 0.9 mL, 1.4 mL, and 2.0 mL) of trisodium citrate solution was added and kept for stirring for about 15 minutes to obtain ruby red, 10 nm AuNPs; the color of the mixture first changed to colorless, then to ruby red for different-sized AuNPs. Then it was cooled to room temperature, and then it was centrifuged and washed with DI water to obtain citrate-coated 10, 30, 50, and 70 nm AuNPs. We measured the average hydrodynamic diameters of the gold nanoparticles by DLS (Malvern Zetasizer Nano ZS). The instrument utilizes a He–Ne laser (4 mW) with a wavelength of 633 nm. The detection angle is 173°, a configuration termed backscatter detection. All samples for DLS measurements were prepared in a 10 mM NaCl aqueous solution [26,27,28].

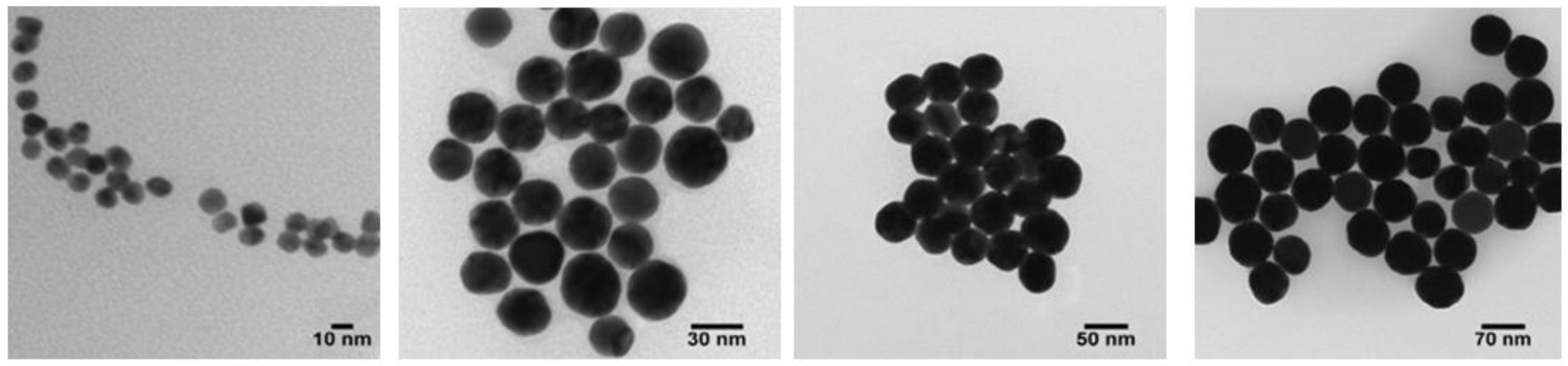

a) TEM images of gold nanoparticles

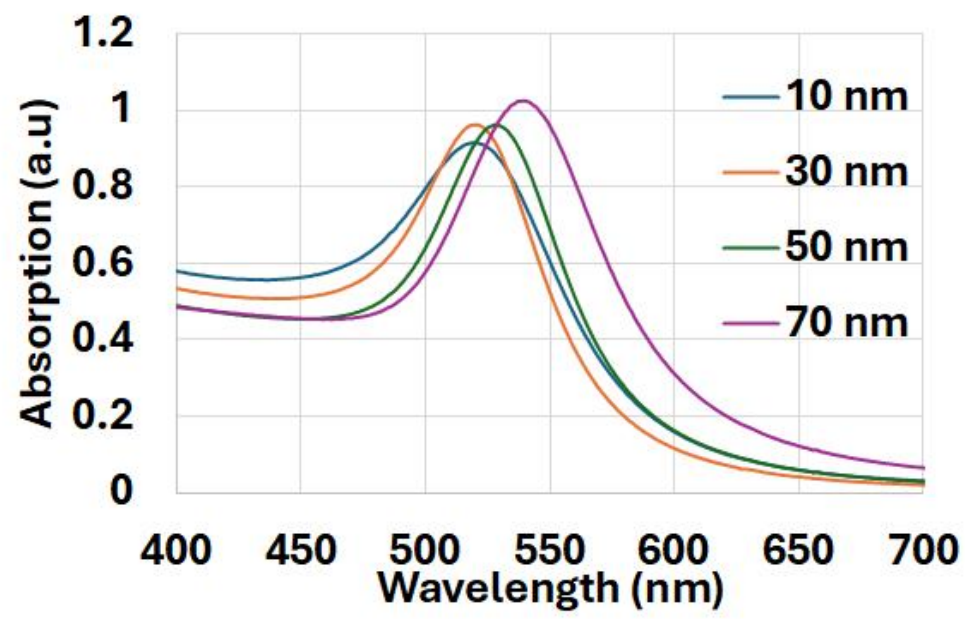

b) Experimental UV-Vis spectra

*Figure.1. a) TEM images of gold nanoparticles and b) Experimental UV-Vis spectra*

## 2.2 Robust PCA

Robust PCA (RPCA) generalizes Principal Component Analysis (PCA) for data containing gross corruption and outliers. Applications of RPCA include computer vision (background modeling and subtraction), anomaly detection, recommender systems, etc., all applications where real-world datasets are often "messy" and contain a significant percentage of outliers [29]. Unlike PCA, which is very sensitive to noise and outliers (it minimizes squared error), RPCA splits the data matrix D into two components:

$$\mathbf{D=L+S} \qquad \textbf{Eq.1}$$

where L is a low-rank matrix that captures the underlying structure, and S is a sparse matrix that represents outliers or noise. The decomposition is typically formulated as an optimization problem:

$$\min_{L,S} ||\boldsymbol{L}||_* + \boldsymbol{\lambda}||\boldsymbol{S}||_1 \text{ subject to } \mathbf{D=L+S} \qquad \textbf{Eq.2}$$

Here, $||\boldsymbol{L}||_*$ is the nuclear norm (sum of singular values) promoting low-rankness, and $||\boldsymbol{S}||_1$ is the L1-norm promoting sparsity. The parameter λ balances the two terms. In the context of image analysis, RPCA can separate the main content (e.g., background) from sparse anomalies (e.g., text, shadows, or noise), as demonstrated in the code above.

## 2.3 Singular Value Decomposition (SVD)

SVD is a fundamental matrix factorization technique. For any real matrix A, SVD expresses it as:

$$\mathbf{A=U\Sigma V^T} \qquad \textbf{Eq.3}$$

where U and V are orthogonal matrices, and Σ is a diagonal matrix of singular values. SVD is also an important step in PCA and RPCA. The SVD algorithm gives us the optimal low-rank approximation of a matrix in the least-squares sense [30,31]. In the code, after performing RPCA decomposition, we applied SVD to the low-rank and sparse matrices to further examine their structure. The left singular vectors (U) and right singular vectors (V) of a low-rank matrix represent orthogonal directions in the data space (the sparse matrix), and the singular values (Σ) indicate the significance of each direction. Figure 2 illustrates TEM images of GNP chains and their low-rank and sparse decompositions obtained by RPCA.

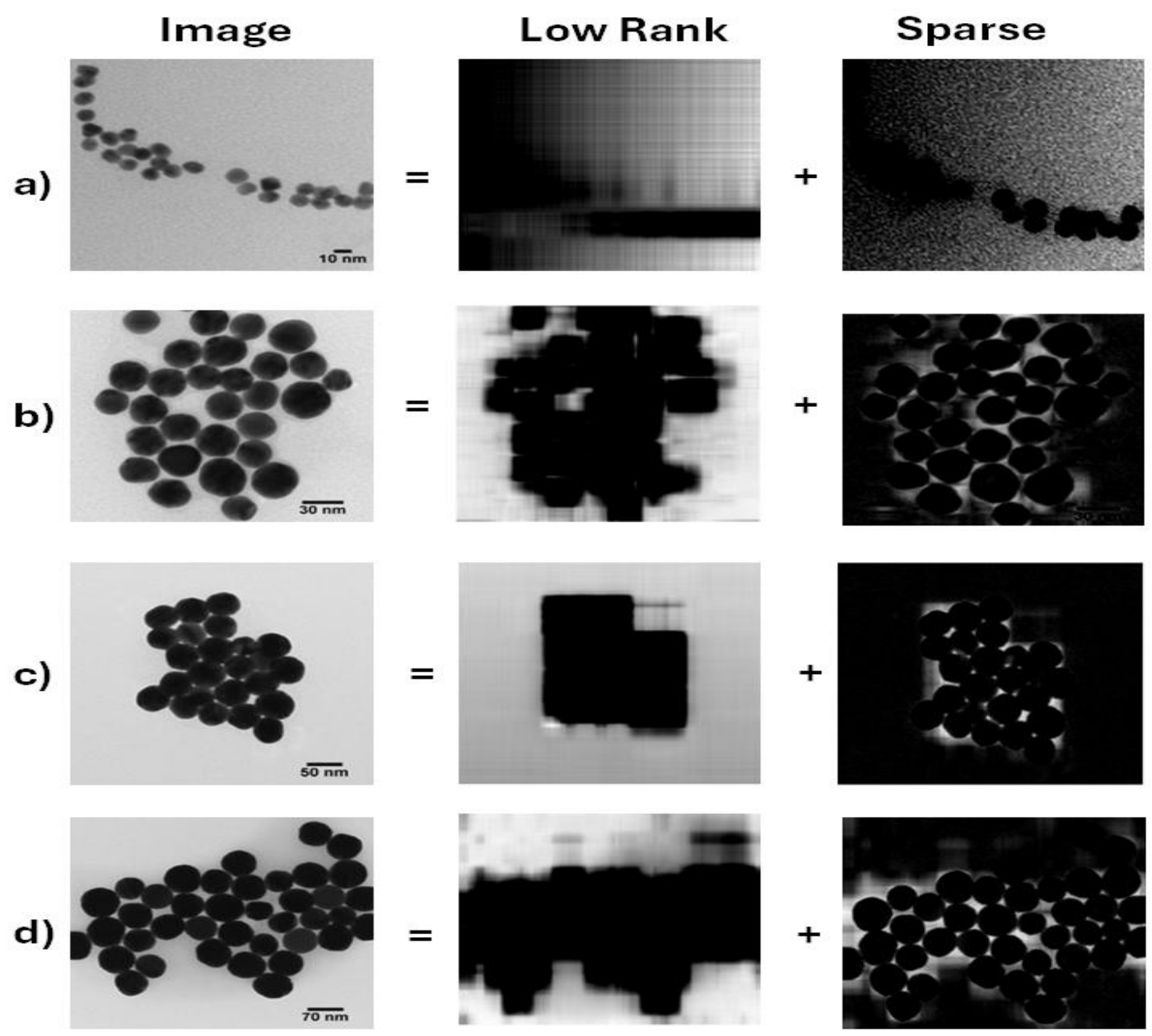

*Figure 2. a) TEM image of 10 nm and its low-rank and sparse decomposition, b) TEM image of 30 nm and its low-rank and sparse decomposition, c) TEM image of 50 nm and its low-rank and sparse decomposition, d) TEM image of 70 nm and its low-rank and sparse decomposition images*

## 2.4 Connection to Hilbert Spaces

A Hilbert space extends Euclidean space to infinite-dimensional complete inner-product spaces. Orthogonality, projection, and decomposition therefore carry over: PCA and SVD are projections of a datum (e.g., an image) onto an orthonormal basis of principal or singular vectors. Under the low-rank assumption, RPCA is projection onto a low-dimensional Hilbert subspace that retains most of the signal while isolating sparse corruption, a best-approximation problem in a prescribed function space [32, 33].

Hilbert spaces also suit the geometry of smooth knots. Physical knots realized as ordered nanoparticle assemblies that host surface-plasmon polaritons (SPPs) may be treated as embeddings $\boldsymbol{\gamma}: \boldsymbol{S^1} \rightarrow \mathbb{R}^3$in a Sobolev space such as $\boldsymbol{H^2}(\boldsymbol{S^1}, \mathbb{R}^3)$. Inner products, completeness, and spectral decomposition then support variational knot-energy analysis and orthogonal reduction. The 3-D plots of $|\,\mathbf{PC1}\rangle$, $|\,\mathbf{PC2}\rangle$, and $|\,\mathbf{PC3}\rangle$for the low-rank and sparse components in Figure 3 can be read in this sense as knotted filaments in feature space [34].

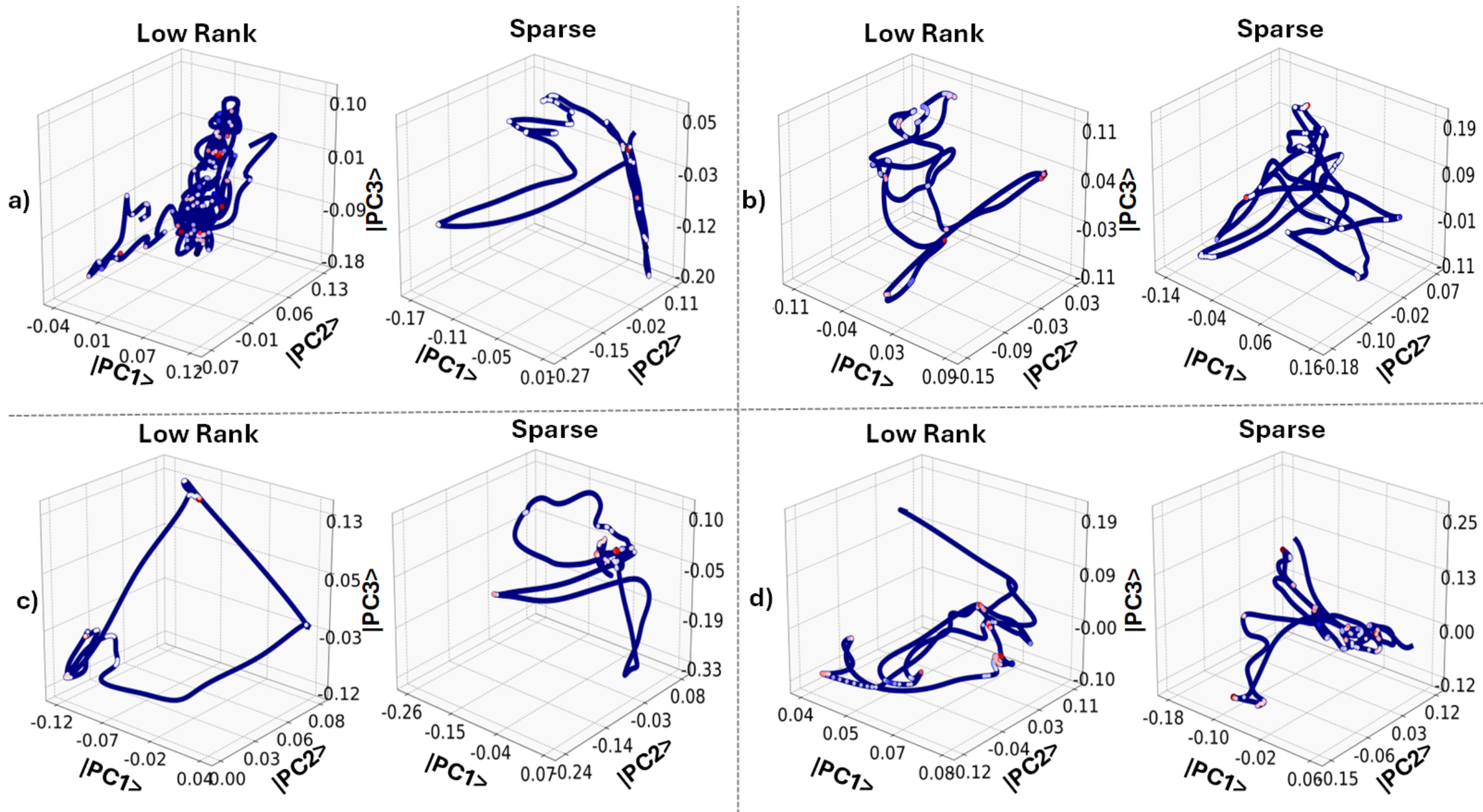

*Figure 3. a) 3D PC plots of low-rank and sparse decomposition of 10 nm; b) 3D PC plots of low-rank and sparse decomposition of 30 nm; c) 3D PC plots of low-rank and sparse decomposition of 50 nm; d) 3D PC plots of low-rank(static) and sparse(dynamic) decomposition of 70 nm.*

Ordered gold-nanosphere strings can sustain SPPs along the chain; curvature and inter-particle coupling can stabilize local plasmonic modes with nontrivial spatial organization [35, 36]. However, the apparently knot-like phase-space loops of PC filaments may not be physical. In bright-field TEM, elastic electron scattering from high-Z gold could produce contrast arising from differences in material mass/thickness and from diffraction effects, including overlaps, aggregates, thickness variations, boundaries, and background from the supporting materials. Inelastic electron scattering could also excite plasmons in various regions. When the electron beam causes charging, contamination, crystal defects, and damage, it can also drive diffusion, rearrangement, and even coalescence. If the PCs are just a small part of a single image and do not change with time, then these loops show spatial organization, not nanoparticle motion over time [16–22].

Thus, the knot-like phase-space loops are inferred from transformed coordinates; they do not appear as filaments in physical space [37]. 'Knotted' means that a closed loop appears to fold itself; it does not in itself imply a nontrivial knot mathematically. Cosmic strings, flux loops and spin networks also appear to be knotted but are not classical knots [24, 38–40]. Nevertheless, a genuine knot requires the existence of an ambient 3-manifold $M^3$, an embedding $S^1 \rightarrow M^3$ which allows no self-intersections, and stable invariants[6].In the RPCA analysis, the feature's persistence and strong closure at the endpoint warrant treatment of the PC1–PC2–PC3 loop as a robust, closed, and apparently knot-like phase-space loop in feature space.

## 3. Theoretical Background

This study uses two theoretical languages, each applying to different objects. Knot invariants classify a closed curve up to topology and geometry (type, writhe, twist, self-linking). Vortex dynamics, including the Moffatt–Ricca helicity of a framed filament, describes rotational structure and its evolution in a physical field. Together they are a coordinating program: invariants for the identity of a closed loop, vortex/helicity language for how a filament could evolve.

In the present TEM/RPCA analysis, the closed object is a knot-like loop in PC1–PC2–PC3 feature space. Knot invariants computed here are those of the score embedding. Vortex dynamics is the intended comparison theory once a map from the image matrix to a framed curve in the specimen (tomography, EELS/CL, or Maxwell/BEM) exists. The subsections below recall the elements of each language that we actually use: classification of feature-space loops, and the Călugăreanu–White–Fuller/helicity pairing as a prospective link to plasmonic vortices, not as a demonstration that the two coincide.

### 3.1 Knot Invariants

A knot invariant is a quantity associated with the knot that does not change under any continuous deformation (an ambient isotopy) of the knot. It captures the knot type in a precise topological sense, regardless of embedding [41]. Knot invariants for knotted optical fields ideally characterize their topology quantitatively. The current work uses six principal invariants, which are briefly described here.

**3.1.1 Crossing Number:** The crossing number $c(K)$ is the minimum number of crossings obtained over all possible knot projections in a plane and is the simplest knot complexity measure available. It uniquely identifies unambiguous knots among the others (simple knots, those with $c \leq 7$) and helps rank all knots by their topological complexity. In physical field systems, the crossing number of a knotted line indicates the complexity of the field pattern and reflects the amplitude and structure of scalar-field gradients in space.

**3.1.2 Ropelength:** The ropelength $L(K)$ of a knot is a ratio of the arc length over the tubular radius (the thickness), where it is used to describe how tightly a knot can be tied with a rope with a constant diameter. Formally,

$$L(K) = \mathrm{len}(K) / \rho(K) \qquad \text{Eq.5}$$

where $\rho(K)$ is the thickness of $K$. Rope length provides geometric information independent of topology, meaning knots with the same topology may have substantially different rope lengths depending on their geometric placement[42,43].

**3.1.3 Alexander Polynomial:** The Alexander polynomial $\Delta(t)$ is a Laurent polynomial in $t$ and is a topological invariant of oriented knots, obtained from the homology of the infinite cyclic cover of the knot complement [25]. It is the first polynomial knot invariant discovered and, because of its simplicity, one of the most widely used. The Alexander polynomial for the trefoil knot $T(2,3)$, the simplest non-trivial torus knot, is given as

$$\Delta(t) = 1 - t + t^2 \qquad \text{Eq.6}$$

for which the knot determinant $|\Delta(-1)| = 3$. Notice that we express all score values in terms of the Alexander polynomial determinant of some reference knot. For instance, notice that open or multi-component curves for which the closed-curve polynomial is not defined will have score values of less than 1 (the unknot has a score of exactly 1). Thus, the Alexander polynomial distinguishes the trefoil knot from the unknot $(\Delta = 1)$ and from more complex knots. However, the Alexander polynomial does not discriminate mirror image knots [44].

**3.1.4 Twist, Writhe, and the Călugăreanu–White Theorem:** For a closed smooth and regular ribbon-like curve in three-dimensional space, the twisting $Tw$ and writhe $Wr$ represent the overall angle of rotation of the normal of the ribbon with respect to its tangent along the curve and the averaged signed number of crossings of the curve with itself in all of its projections into planes, respectively, and also characterize the non-planarity or supercoiling of the curve in general [45,46]. These two geometric parameters are embedded in the self-linking number $L_k$ topological invariant for closed curves, and are mandated by the Călugăreanu – White – Fuller theorem:

$$L_k = Tw_r + W_r \quad \textbf{Eq.7}$$

This fundamental equation defines the integer-valued topological invariant $L_k$ as the sum of the local and global factors. The local factor, which relies on the embedding, and the global factor are constant topological invariants (Figure 6). $T_{wr}$ captures the local helix winding of the field, $Wr$ the global chirality, and $L_k$ the self-entanglement. The twists derived from the torsion proxy, which depends on the differential geometry of Frenet-Serret, are known as $T_{wr}$. The signed contributions are accumulated by adding two pairs of segments of the whole segments. Therefore, the sign denotes the global coiling handedness. Positive local torsional winding is indicated as $T_{wr} > 0$, and negative local torsional winding as $T_{wr} < 0$.

We use the Gauss writhe model to study the curve's 3D coiling. This model calculates the amount and handedness of global 3D coiling in the curve from each pair of segments in the curve. If the writhe value is positive $(Wr > 0)$, the curve possesses positive or right-handed coiling. When it is negative $(Wr < 0)$, it shows negative, or left-handed, coiling. If the absolute value of the writhe value $(|Wr|)$ is large, the curve exhibits strong global 3D coiling [47,48].

**3.2.5 Mean Curvature and Mean Torsion:** The curvature $\kappa(s)$ and torsion $\tau(s)$ of a space curve are the classic differential-geometric invariants representing its finite deformations as a function of arc length, described through the corresponding Frenet-Serret equations.

$$dT/ds = \kappa N, \quad \textbf{Eq.8}$$

$$dN/ds = -\kappa T + \tau B \quad \textbf{Eq.9}$$

$$dB/ds = -\tau N \quad \textbf{Eq.10}$$

Here, $T$, $N$, **and** $B$ are the unit tangent, normal, and binormal vectors. Curvature $\kappa$ characterizes the deviation of a curve from the tangent direction, which has unit length specified by the vector $\langle\tau\rangle$; torsion $\tau$ indicates a turning of the osculating plane along the tangent vector; it quantifies the twisting of the osculating plane around the curve. The curvatures and torsions averaged over the knot are denoted by $\langle\kappa\rangle$ and $\langle\tau\rangle$, respectively; their mean values characterize the spatial variations of curvature and torsion along a unit length of the knotted curve. The sign of the mean value of torsion carries significant information about the spiral structure of the knot; a positive value for $\langle\tau\rangle$ indicates a right-handed helicity (i.e., the direction of the rotation of the binormal vector relative to the tangent knot vector $\langle \boldsymbol{T} \rangle$ matches a right-handed helicity), and a negative value indicates a left-handed helicity [49].

### 3.2 Vortex Dynamics in Optical and Plasmonic Fields

The light propagating in nanophotonic systems can form vortices: loci where the amplitude of the field is zero and its phase winds by $\boldsymbol{2\pi m (m} \in \mathbb{Z}\boldsymbol{)}$**.** The phase singularities are structurally similar to the vortices in a rotational flow [47,48]. Singular optics borrowed ideas of helicity, circulation, and reconnection from fluid mechanics [24,49,50]. The only difference is that the Helmholtz circulation statement concerns a material vorticity field, whereas an optical vortex is a singularity of a complex amplitude (or polarization) [51].

That analogy is not automatic on the level of integrals. The Moffatt–Ricca helicity [24,49] is defined for a divergence-free vector field (vorticity, or a magnetic flux tube) and, via the Călugăreanu–White–Fuller relation, for a framed filament in $\mathbb{R}^3$. An optical or plasmonic vortex is characterized first by its topological charge (and, for vector fields, by polarization singularities [51]). Equating the two quantities requires identifying a concrete filament, e.g., a phase isophase, a Poynting-vector line, or an E-line, and stating the framing. The plasmonic near field does support vortices (orbital angular momentum of SPPs, phase singularities at gaps or around particles), but only in the electromagnetic field itself.

Vortex dynamics provides a complementary yet equivalent theoretical basis for studying knotted field configurations. In the particular case of optical fields and electromagnetic fields, a vortex is a phase singularity. It is a location or set of locations where the field amplitude tends to zero, and the phase is therefore indeterminate. Around these points, the field phase winds by an integer multiple of $\boldsymbol{2\pi}$ [47,51]. These optical vortices relate to vortex filaments in fluid dynamics as electromagnetic fields do, and much of the theoretical work on fluid vortex dynamics is directly transferable to the optical field. This work does not claim that the knot-like loop of a closed object extracted from TEM/RPCA is an optical or plasmonic vortex; it is a comparative theory for a future field-based curve (EELS, CL, or Maxwell/BEM). The paper claims static and dynamic PC loops may encode plasmonically relevant morphology; it is not a claim about reconstructing phase singularities of $\boldsymbol{E}(\boldsymbol{r})$ using score embedding.

**3.2.1 Moffatt–Ricca Helicity:** Helicity in the vortex-dynamics sense was introduced by Moffatt (1969) as the volume integral of the helicity density, the inner product of a solenoidal vector field u with its [24]

$$\mathbf{curl}\ \boldsymbol{\omega} = \nabla \times \mathbf{u}:\ \mathbf{H} = \int \mathbf{u} \cdot \boldsymbol{\omega}\ \mathbf{dV} \qquad \textbf{Eq.11}$$

In ideal (inviscid, non-diffusive) flow, H is a conserved topological invariant that measures the degree of linkage and knotting of the vortex lines of u. Ricca and Moffatt (1992) subsequently demonstrated that for a thin knotted vortex filament, the helicity decomposes exactly as

$$\mathbf{H} = \boldsymbol{\Gamma}^2\ (\mathbf{Wr} + \mathbf{Tw}) = \boldsymbol{\Gamma}^2\ \mathbf{L_k} \qquad \textbf{Eq.12}$$

where $\Gamma$ is the circulation of the vortex [25]. Moffatt–Ricca helicity is linked to the self-linking number of knots. It naturally connects vortex dynamics to knot invariants.

Additional work by Ricca (2008) showed that helicity provided a lower bound on the energy of knotted and linked flux tubes under zero framing (i.e., no net linking in the ribbon framing of the knots/links) [49,52]. To avoid future confusion, we stress that the quantity $H \approx \Gamma^2 SL$ is the Moffatt–Ricca self-helicity contribution of a single filament and not the general total helicity. For the multi-component links, we include a mutual helicity $\mathbf{2 \sum_{i<j} \Gamma_i \Gamma_j Lk_{ij}}$, in which every link component carries its own circulation ˜***Ci***. This result also corresponds to an energy topology inequality. The sign of H shows the handedness of the dominant vortex structure: $H > 0$ stands for right-handed ($L_k > 0$) vortex self-linking. $H < 0$ corresponds to left-handed self-linking ($L_k < 0$).

**3.2.2 Kelvin Spectral Quality:** The Kelvin spectral quality (KSQ) is a dimensionless measure derived from spectral analysis of the vorticity field, quantifying the degree to which the vortex energy is spectrally concentrated in sharply defined and spatially coherent vortex filaments as opposed to a broad and disordered spectrum of vortical features [48]. A high KSQ value indicates sharp, well-defined, spectrally coherent vortex filaments, as expected in a self-organized or resonantly driven state, whereas a low KSQ indicates diffuse, disordered vorticity. KSQ relates to Kelvin's classical circulation theorem [53] by measuring the spectral sharpness of the circulation distribution and serving as a sensitive diagnostic of the vortex system's dynamical state and phase coherence.

**3.2.3 Vortex Reconnection and Minimum Distance:** The phenomenon of vortex reconnection describes the topological behavior in which two vortex filaments, approaching each other at a distance lower than some critical diameter, change their "topological connection" and the topology of the entire vortex system [54]. Under ideal conditions, vortex-filament topology is preserved (see Kelvin's theorem); however, in dissipative or forced systems, reconnections enable topological change and are among the leading mechanisms for the appearance and disappearance of knotted vortex structures. Kleckner and Irvine (2013) experimentally demonstrated reconnection in knotted hydrodynamic vortices [55]. The reconnection distance $\delta_{min}$ denotes the minimum distance between different vortex filaments just before reconnection. Reconnection events can occur only at small $\delta_{min}$ values, indicating high local field gradients and a high probability of topological change. Based on simulations by Laing, Ricca, and Sumners (2015), writhe helicity is conserved for anti-parallel reconnection, where the tangents on the two approaching strands are opposite; it is not conserved for any other reconnection geometry. Writhe helicity conservation further supports the links between reconnection dynamics and invariants of knot theory discussed above [56].

**3.2.4 Composite Topological Entanglement Index:** To diagnose topological entanglement complexity in a heuristic sense, a composite topological-entanglement index (CTEI) that encompasses normalized metrics of filament proximity, curvature, self-approach, and torsional complexity is introduced. The Composite Topological Entanglement Index (CTEI) is a geometric measure that draws on topology to provide an overall measure of filament complexity. It combines normalized crossing density, curvature, self-approach, and torsion, similar to the work of Moffatt and Ricca, where they defined vortex helicity and geometric arrangement [24]:

$$\mathbf{CTEI = w_1C_d + w_2K + w_3S_a + w_4T} \qquad \textbf{Eq.14}$$

where $C_d$ is normalized close-contact/minimum distance, $K$ is normalized curvature or bending complexity, Sa is a normalized self-approach/spatial entanglement measure, T is normalized torsional/twist complexity, and ***w1,..,w4*** are user-selected weights such that ***w1+w2+w3+w4=1***.

We kept the weights constant across all datasets and perturbed them to test sensitivity to individual weights, such that $\sum_i w_i = 1$. We assigned these weights to the CTEI based on the relative importance of each component in describing entanglement. Knot/topological complexity (w2=0.35) makes the largest contribution to the CTEI because knot theory directly quantifies entanglement. Because this property is important and static (w4=0.30), we assign the next-highest weight to topological persistence. Spatial organization (w3=0.20) captures the complexity of the 3D structure, and directional coherence (w1=0.15) has the lowest weight because it measures loop smoothness. An equal-weight scheme ($w_i$ = 0.25) also yields the same trend but slightly different observations. These weights are physically motivated and not universal constants. Higher CTEI values indicate greater three-dimensional entanglement and geometric complexity. Like the Coherence Score, CTEI is a continuous geometric diagnostic that measures the degree of filament winding and self-approaching, regardless of knot type. From the Alexander polynomial test of Section 4.1, we see that CTEI is also different from topological entanglement.

## 4. Results and Discussion

The two frameworks described in chapter 3 here are not entirely independent. The Călugăreanu–White–Fuller theorem and the Moffatt–Ricca helicity decomposition establish a mathematical mapping between the kinematical-dynamical quantities associated with a knot-like curve in the differential geometry of curves ($Tw, Wr, \kappa, \tau$), and the helicity-curl dynamical-physical quantities associated with the vortex (H, vortex energy, reconnection behavior). The self-linking number $L_k$, a topological invariant of a knotted loop, is proportional to the normalized Moffatt–Ricca helicity $H/\Gamma^2$; the quantities $T_w$ *and* $W_r$ are the geometric descriptors that correspond to the decompositions of $L_k$, $H$; the mean torsion $<\tau>$ of the curve represents the local phase winding rate, which is responsible for generating reconnection events of the vortex [23,24,25]. The spectral quality $H$ of the Kelvin waves and the minimum reconnection distance provide dynamical information about the system's ability to dynamically conserve topologically distinct states. We use this framework throughout Section 4 to interpret the results for knot invariants and vortex diagnostics within the problem's underlying topology-geometry-physics framework.

### 4.1 Crossing Number, Rope Length and Alexander Polynomial

Figure 4 presents the crossing number, rope length, and Alexander-derived score for both static and dynamic loops extracted from the GNP TEM/RPCA analysis. For both categories, we obtain the crossing number at every diameter. The static case at 70 nm has the lowest crossing number, 10. Dynamic crossings are almost 3-4 times higher than static crossings. Rope length falls with diameter for both classes (the extracted curve is less elongated relative to its thickness); the dynamic rope length is about 2× the static value, so the dynamic curves are longer and/or thinner in the same embedding [43]. The Alexander determinant is 1 for both classes except at two sizes: 3 for the static loop at 10 nm, and 3 for the dynamic loop at 30 nm. Those two curves are not Alexander-trivial. All other closed loops have determinant 1, so their crossings are canceled. A determinant of 3 rules out the unknot; it does not, by itself, name the knot. This contrasts with the stretched, more tangled knot-like loops called dynamic ones.

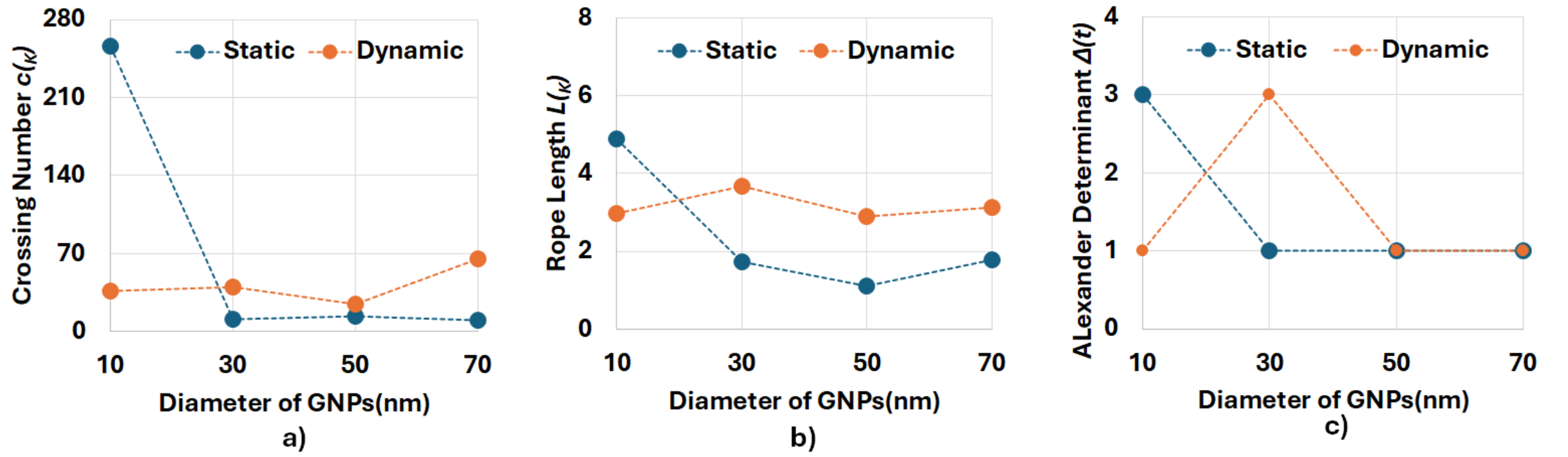

*Figure 4. a) Crossing number, b) rope length, and c) Alexander Indicator of the static and dynamic cases*

### 4.2 Twist, Writhe and Self-Linking Number

Figure 5 shows plots of twist, writhe, and self-linking for the static and dynamic configurations. The static configuration has a positive writhe trend, while the dynamic writhe is negative (Figure 5a). The opposite trend in writhe indicates opposite global coiling directions (i.e., opposite handedness) and different geometric organization between these configurations in phase space. Meanwhile, the torsion-derived twist, $\boldsymbol{T_{w\tau} = (1/2\pi) \oint C\ \tau(s)\ ds}$, changes from positive at 10 and 30 nm to negative at 50 and 70 nm, revealing a diameter-dependent reversal in local handedness (Figure 5b). Consequently, the geometric self-linking proxy, $\boldsymbol{SL_\tau = W_r + T_{w\tau}}$, decreases toward ~1.0 for dynamic and −1.1 for static, respectively, for the static and dynamic fluctuations (Figure 5c) and shows a different net balance between global coiling and local twist [24,45].

The observed diameter-dependent transition from positive to negative torsion-derived twist may indirectly reflect size-dependent electron-beam–nanoparticle interactions. Increasing GNP diameter modifies projected thickness, elastic and inelastic scattering, plasmon excitation, energy deposition, and particle–support coupling, thereby changing the spatial contrast patterns encoded by the principal components. These changes could alter global coiling, local torsional handedness, and self-linking proxy of the resulting feature-space loops. However, because these quantities are derived from PC1–PC2–PC3 image-feature coordinates, they should not be interpreted as direct physical rotation or vortex motion of the nanoparticles. Establishing a beam-induced origin would require dose-resolved TEM measurements, low-dose controls, or spatially resolved EELS [17,18,19].

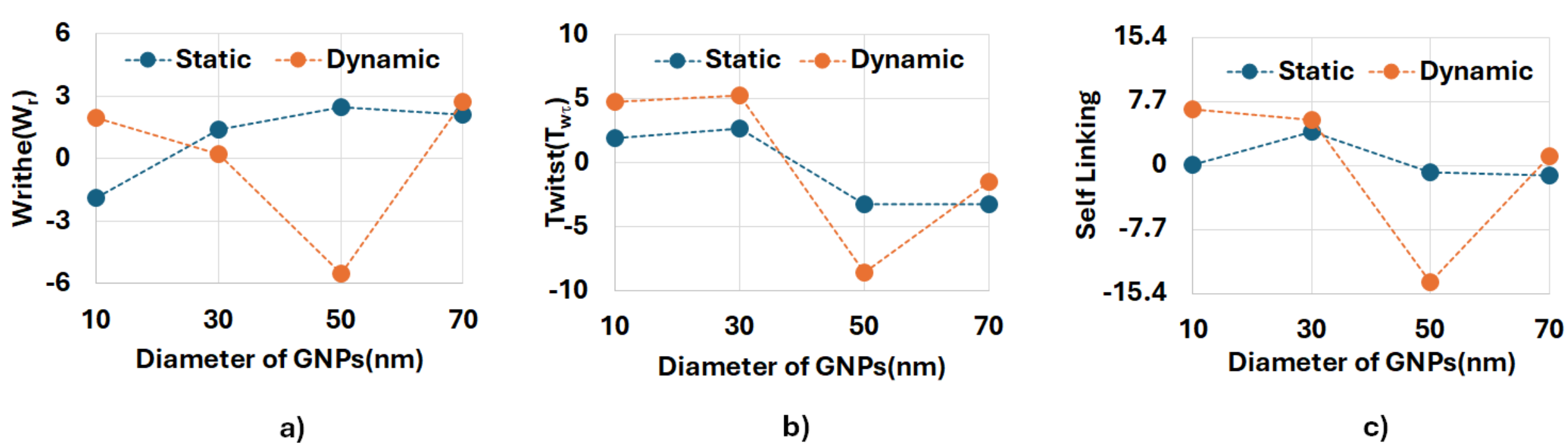

*Figure 5. a) Twist, b) writhe, and c) self-linking calculation of the static and dynamic cases*

### 4.3 Mean Curvature and Mean Torsion

Figures 6a and 6b show simulations of mean curvature $\langle\kappa\rangle$ and mean torsion $\langle\tau\rangle$ for static and dynamic configurations, correspondingly. These graphs show an almost inverse relationship between curvature and torsion: when curvature was significant, loops were flatter; lower curvature was associated with higher torsional twisting (Figure 3c, low rank), and higher twisting was associated with less bending (Figure 2b, sparse). In general, the mean curvature of static cases is much larger than that of dynamic loops. The larger mean curvature means that the static case has the more bent shared score path of the two, even if the dynamic case is longer and is crossed about twice as much.

More interestingly, the mean torsion drops with diameter for both categories and is negative for the largest diameters. This drop in $\langle\tau\rangle$ means that the bending plane rotates more slowly when the projected particle gets larger. At the upper end, where the mean torsion is negative, the rotation reverses. This parallels the twist $(T_{w\tau})$ trends, where the sign of the torsion-derived twist already flips inside both blocks. This is a change in the local loop geometry [49]. This behavior may relate to electron-scattering contrast, electronic excitations, and spatial correlations between electrons and nanoparticles in TEM. This can affect the data's covariance structure, which further alters the torsional geometry of the constructed phase space [17,18,19].

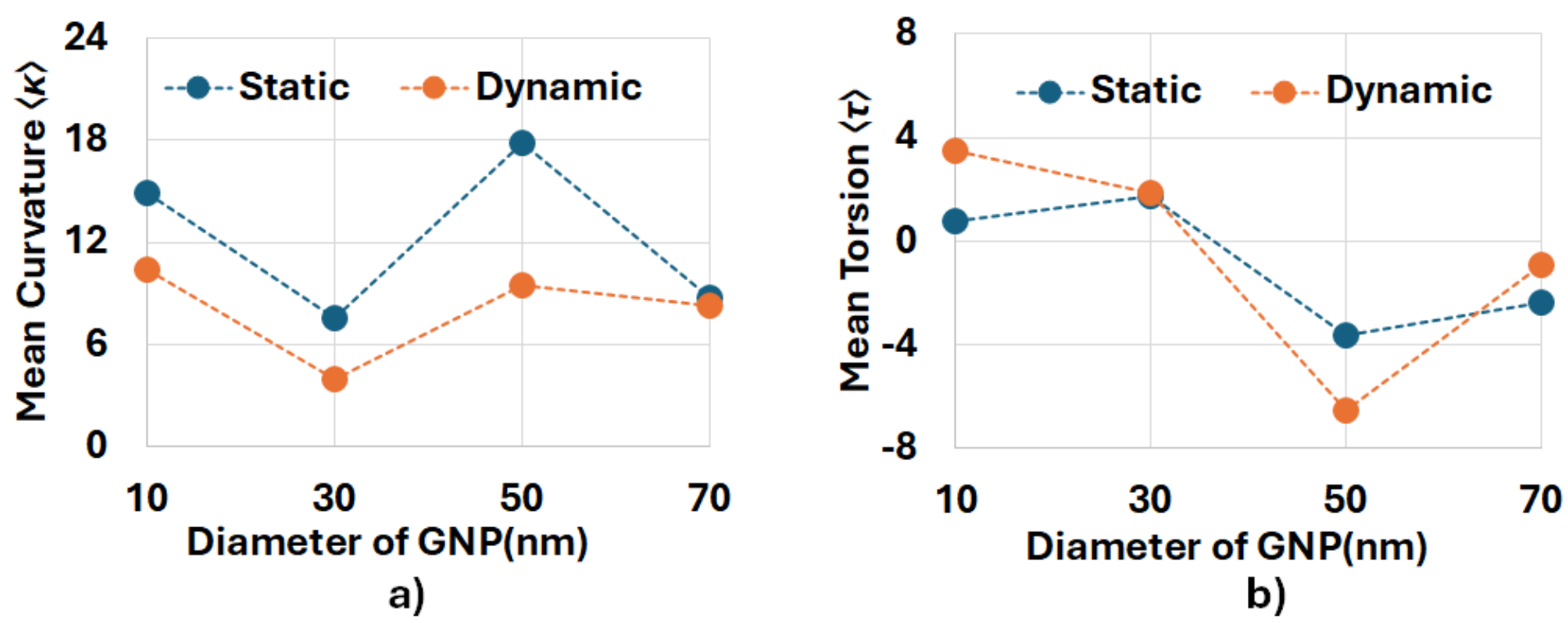

*Figure 6. a) Mean Curvature and b) Mean Torsion calculation of the static and dynamic cases*

### 4.4 Moffatt–Ricca Helicity and Vortex Energy

Figure 7 presents plots of the Moffatt- Ricca helicity for the static and dynamic configurations. Similar trends are noticed between the helicity (Moffatt–Ricca Helicities) and twist, but the trend has a different scale (only around $10^{-14}$). This depends on the flow circulation, which dictates its helicity. Because the writhe calculation yields global twisting differences between the two geometries (opposite coiling orientations), the self-linking decreases toward −1 and +1 for the static and dynamic systems, respectively. This gives an idea of the amount of coiling and twisting present in both geometries in phase space [57,58].

In terms of diameter dependence, the Moffatt–Ricca helicity is almost independent of diameter in both regimes. The positive Moffatt–Ricca helicity in the dynamic configuration implies net right-handed self-linking of the vortex filament, consistent with the positive twist (Tw) and near-zero writhe ($W_r$) observed at large diameters (Section 4.2). In the dynamic case, twist dominates ribbon helicity, providing a coherently positive contribution [23]. Conversely, the strongly negative

Moffatt–Ricca helicity of the static configurations reflects the dominance of a negative twist ***($T_w$)*** at all diameters, again consistent with the knot-invariant analysis. The substantial difference between the static and dynamic helicity values highlights how dynamic excitation changes the sign of self-linking and the qualitative form of the vortex filament. The coherent phase change of the electron beam-induced plasmonic oscillation during driving ensures that twist and writhe lock into a configuration of high positive twist with very low writhe [47]. Despite these promising findings, the geometric properties of the feature-space loop, although reminiscent of a superfluid vortex filament, do not demonstrate that the physical superfluid has helicity.

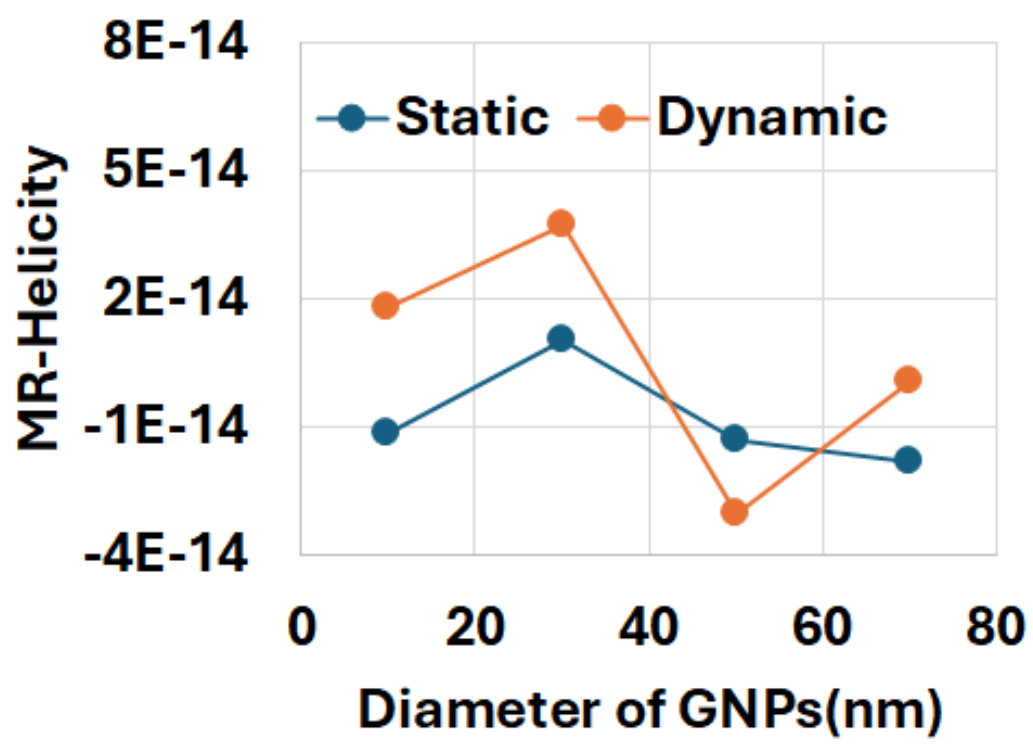

*Figure 7. Moffatt-Ricca Helicity of the static and dynamic cases*

### 4.5 Kelvin Spectral Quality and Reconnection Minimum Distance

Kelvin spectral quality (KSQ) is a dimensionless quantity that characterizes the spectrally coherent structure of the vorticity and its spatial organization in the flow field; large KSQ suggests coherent vortex filaments, while low KSQ suggests diffuse vorticity fields [48]. The data indicate that spectral coherence of the dynamic configuration progressively increased with the diameter of the GNP. A planar path in phase space characterizes flow with a well-defined length scale. This behavior indicates a shift from globally entangled three-dimensional dynamics to a more scale-selected, mode-dominated regime. The Kelvin-like spectral signature is then enhanced. In contrast, the static configuration retains a broad spatial spectrum that shows multiscale geometrical organization over several characteristic length scales.

Vortex re-connection, the process by which vortex lines swap topology when they close and cross, is a major dynamic process shaping knotted loop configurations [54,55]. The re-connection minimum distance, measured as the smallest separation between two distinct vortex filaments before re-connection, was 0.0008 and 0.008 for dynamic and static configurations, respectively, a 10-fold reduction in the dynamic case, where these phase-space curves approached each other more closely. Furthermore, the minimum reconnection distance in the dynamic loop is about an order of magnitude smaller. This implies a transition from a multiscale 3D configuration to more planar dynamics with a localized spectrum and local interactions.

The TEM does not physically bring two vortex filaments to 0.0008 on the particle. It just changed the image. The dynamic configurations in phase space record this change. The static curve is the silhouette of the same GNP (shape, thickness, gaps) whose closest self-approach (0.008) is about ten times larger because it is moving slower. The beam-particle interaction that caused the charging, a hop or rotation on the carbon, a contamination spot that changes the beam-particle interaction, or two particles that happen to be stacked on top of each other, causes a quick contrast change in a short burst of TEM frames. The sparse score causes the system to swing back into previously visited regions, bringing non-adjacent parts of the dynamic configurations much closer together (0.0008). Therefore, the minimum distance can be shorter than ten times. Besides, those ones have more crossings and longer ropes [16,17,18, 19].

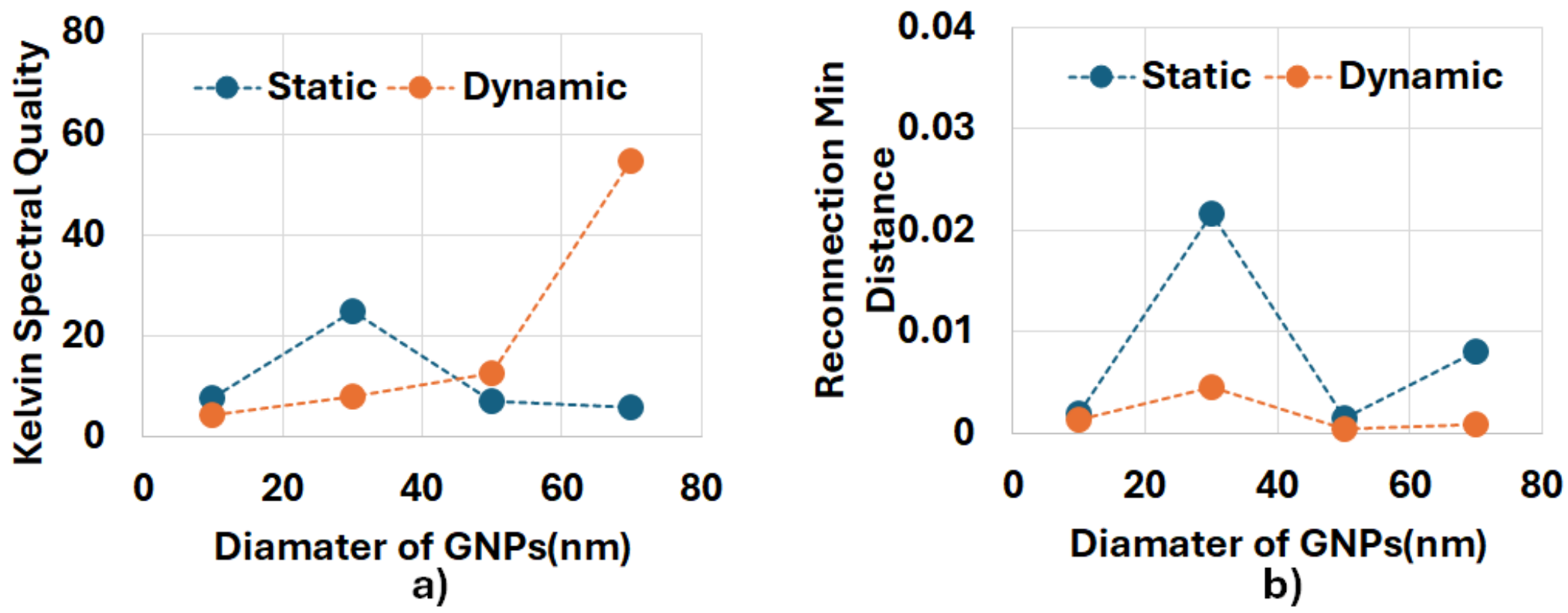

*Figure 8. a) Kelvin Spectral Quality and b) Reconnection Minimum Distance of the static and dynamic cases*

### 4.6 Composite Topological Entanglement Index (CTEI)

The PC1-PC2-PC3 loops in Figure 9 showed a decrease in CTEI from 0.9 to 0.6 as GNP diameter increased under both static and dynamic settings. The results suggest that increasing GNP size reduced loop entanglement, yet the loops remained topologically organized and highly entangled (0.45, 0.5) in phase space, as indicated by CTEI. Another factor contributing to the TEM analysis was nanoparticle size. As size increases, more electrons scatter, and entanglement complexity changes. This phase-space image shows that the CTEI value changes for each phase-space complex entanglement; thus, it decreases as nanoparticle size increases in solution. However, this does not mean the phase-space entanglement disappears; it simply decreases. Figure 8a shows that the Kelvin spectral quality increases with dynamic configurations, and Figure 9a shows that the CTEI decreases as nanoparticle size increases. Surprisingly, this shows a transition from an entangled phase-space organization to spectral localization of Kelvin-like modes [57].

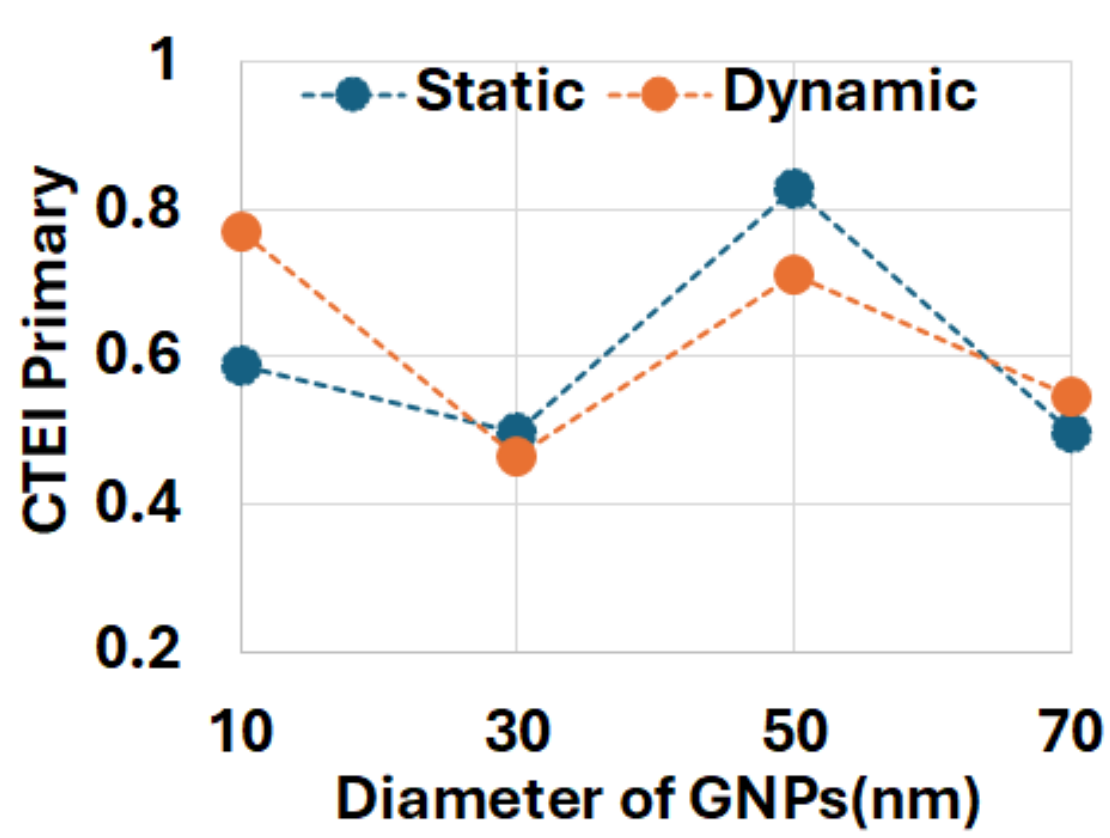

*Figure 9. CTEI of static and dynamic cases*

## 4.7 Lord Kelvin's knotted vortex fluid: Aether

Lord Kelvin (William Thomson) hypothesized in 1867 that the atoms were knotted vortex filaments in the frictionless, all-pervading fluid, called the aether, and, following Helmholtz theorems on the permanence of vorticity in inviscid flow, he suggested that the knots represented the different elements (chemical) based on their distinct topological classification, prompting Tait to pioneer the knots classification [50, 60, 61, 62]. The luminiferous aether was later discovered not to exist but the mathematics that Kelvin set in motion persisted: the topological classification of vortex knots, the conservation of helicity in ideal flow, and the belief that the knot determines the identity of the physical entity all precede the Moffatt-Ricca formalism [24] and the modern-day knot invariant theories [41, 60].

Central among those invariants is the Alexander polynomial $\boldsymbol{\Delta(t)}$, introduced by J. W. Alexander in 1928 [41]. For a genuinely knotted closed curve, a trefoil, for example $\boldsymbol{\Delta(t)}$ is a non-constant Laurent polynomial $(\boldsymbol{\Delta_{trefoil}(t) = 1 - t + t^2});$ for the unknot, and for any curve continuously deformable into it, $\boldsymbol{\Delta(t) = 1}$. The contrast between a topologically nontrivial knot and a geometrically elaborate but trivial curve is a modern one. So the polynomial separates topologically stable knots from topologically ordinary and geometrically difficult-to-describe deformations just as reliably as it separates vortex atoms from ordinary, untangled vortex disturbances. So, the theory has two layers. A stable topological layer says that the knot type is conserved and identifies the element. A geometric/dynamical layer says that the tube's actual shape and motion determine its energy and spectrum[63].

Similarly, our interpretation of reconstructed phase-space structures can be linked to modern Einstein–Aether theories [64,65,66,67]. Einstein–Aether theories incorporate gravitational dynamics and employ a covariant spacetime metric tensor coupled to a dynamical aether field, whereas PCA uses a statistical covariance matrix to identify the principal directions of data variation. Although mathematically distinct, both frameworks support geometrical analysis in their respective spaces. In the present study, we characterize reconstructed knot-like loops using geometric and topological measures, including curvature, torsion, writhe, twist, and knot invariants. The persistent and transient configurations exhibit distinct geometrical characteristics while retaining predominantly similar topological signatures. These findings provide a mathematical analogy to Kelvin's aether as an effective medium that supports geometric deformations without necessarily changing topological identity, independent of gravitational aether theories.

## 5. Summary and Conclusions

Geometrical and topological analysis shows that dynamic crossing and rope length have more crossings than the static case, whereas static phase-space configurations have higher bending and mean curvature. Both classes of phase-space Alexander determinants are 1, but for very small nanoparticle diameters, 3 is present. This implies that if something is geometrically complicated, it is not necessarily topologically complicated. The opposite signs for writhe suggest opposite coiling globally. The opposite twist signs suggest that torsion-derived twist changes from positive to negative as the diameter increases from below 40 nm to above 40 nm. However, the calculated self-linking decreases for both configurations to about -1 and +1, indicating that global coiling and local twisting contribute differently in the two configurations. The opposite twist and writhe signs do not indicate a change in knot topology or in the helicity of the physical vortex.

Mean curvature and mean torsion showed an approximately inverse trend between the static and dynamic configurations, suggesting a redistribution between local bending and out-of-plane rotation in the feature-space loops. This inverse behavior is an empirical characteristic of the present data rather than a universal geometrical relation. Because Frenet torsion can become unstable near low-curvature points, we evaluated the trend using smoothed loops and mean absolute torsion. Interestingly, the torsion-derived twist switches from positive to negative at roughly 40 nm, and mean torsion becomes negative at larger diameters in both the static and dynamic configurations. Mean torsion follows a pattern similar to self-linking; thus, local 3D frame rotation affects writhe and twist.

Due to circulation-dependent scaling in the analogs of Moffatt–Ricca Helicities, they have very small magnitudes, about $10^{-14}$. However, they closely follow the self-linking trend of both static and dynamic configurations. The improved Kelvin-like spectrum, shorter transient self-approach distances, and gradual decline in CTEI suggest that increasing nanoparticle diameter substantially alters the distribution of geometrical and spectral organization. These contrasting trends suggest that geometrical organization is shifting from globally coherent structures toward more localized, mode-specific configurations. Overall results are consistent with Kelvin-inspired geometrical disturbances in the effective medium, possibly induced by the TEM electron beam, but do not support physical vortex reconnection or aether theory. The reconstructed PCA phase space appears to be a mathematical analog of Kelvin's aether: stable and dynamic phases exhibit varying geometrical disturbances while sharing similar knot classifications.

## Acknowledgement

*We thank Nanocomposix Inc. for providing TEM images and nanoparticle spectra. Claude (Anthropic, claude.ai) and ChatGPT (GPT-6) assisted the authors in comparison of theoretical interpretations and manuscript language refinement. The authors are solely responsible for all scientific content, data, results, interpretations, and conclusions, all of which have been independently verified. This research received no specific grant from any funding agency in the public, commercial, or not-for-profit sectors.*

**LIST OF REFERENCES**